\documentclass[preprint,prd,nofootinbib,tightenlines,amsmath]{revtex4}
\usepackage{bbm}
\usepackage{amsfonts}
\usepackage{mathrsfs}
\usepackage{graphicx}
\usepackage{bm}

\begin{document}
\baselineskip=15pt \parskip=5pt

\vspace*{3em}

\title{Study of WIMP annihilations into a pair of on-shell scalar mediators}

\author{Lian-Bao Jia}
\email{jialb@mail.nankai.edu.cn}

\affiliation{School of Science, Southwest University of Science and Technology, Mianyang
621010, P. R. China
}

\begin{abstract}

In this article, we focus on a new scalar $\phi$ mediated scalar/vectorial WIMPs (weakly interacting massive particles) with $\phi$'s mass slightly below the WIMP mass. To explain the Galactic center 1 - 3 GeV gamma-ray excess, here we consider the case that a WIMP pair predominantly annihilates into an on-shell $\phi \phi$ pair with $\phi$ mainly decaying to $\tau \bar{\tau}$. The masses of WIMPs are in a range about 14 - 22 GeV, and the annihilations of WIMPs are phase space suppressed today. In this annihilation scheme, the couplings of the $\phi$ - standard model (SM) particles are almost arbitrary small, and the WIMP-nucleus spin-independent scattering can be tolerant by the present dark matter (DM) direct detections. A scalar mediator-Higgs field mixing is introduced, which is small and available. The lower limit on the couplings of the $\phi$-SM particles set by the thermal equilibrium in the early universe is derived, and this constraint is above the neutrino background for scalar DM in direct detections. The WIMPs may be detectable at the upgraded DM direct detection experiment in the next few years, and the exotic decay $h \rightarrow \phi  \phi$, the production of $\phi$ may be observable at future high-luminosity $e^+ e^-$ collider.

\end{abstract}

\maketitle

\section{Introduction}

The weakly interacting massive particle (WIMP) type dark matter (DM) attracts much attention in DM direct detections, and the cold DM relic density can be derived from thermally freeze-out WIMPs. Today, the compatible confident events are still absent in DM direct detection experiments, and the recent search results of CRESST-II \cite{Angloher:2015ewa}, CDMSlite \cite{Agnese:2015nto}, LUX \cite{Akerib:2015rjg} and XENON1T \cite{Aprile:2015uzo} set stringent constraints on the WIMP-nucleus spin-independent (SI) scattering. Even with these rigorous constraints, the case of the SI interaction being dominant in WIMP-nucleus scattering can still be allowed by the present direct detections, and a feasible scenario will be investigated in this work with the possible DM signatures from indirect detections.

The cosmic ray observations, such as $\gamma$-rays, neutrinos, positrons, and antiprotons from DM dense regions, may indirectly reveal properties of WIMPs. The recent 1-3 GeV gamma-ray excess from the Galactic center may be due to WIMP annihilations, for WIMPs in a mass range about 35-50 GeV annihilating into $b\bar{b}$ with corresponding annihilation cross section $\sim (1-3)\times 10^{-26}$ cm$^3/$s \cite{Goodenough:2009gk,Hooper:2011ti,Abazajian:2014fta,Daylan:2014rsa,Calore:2014xka,Alves:2014yha,Zhou:2014lva}, or WIMPs in a mass range about 7 -11 GeV annihilating into $\tau \bar{\tau}$ with the annihilation cross section $\sim$  $0.5 \times 10^{-26}$ cm$^3/$s (20\% to $b\bar{b}$ also allowed) \cite{Hooper:2010mq,Hooper:2011ti,Abazajian:2014fta,Calore:2014xka,Daylan:2014rsa}. In this work, we focus on the latter case, that is, the main WIMP annihilation products in SM sector are $\tau \bar{\tau}$ pairs (see Refs. \cite{Lacroix:2014eea,Yu:2014mfa,Ibarra:2015fqa,Kim:2015fpa} for more discussions). Moreover, with a small number of visible matter in dwarf satellite galaxies, gamma rays from the DM-dominant dwarf galaxies provide significant information about WIMPs. The $\tau \bar{\tau}$ mode galactic center GeV gamma-ray excess can be compatible with the recent results from the new dwarf spheroidal galaxy observations \cite{Geringer-Sameth:2015lua,Drlica-Wagner:2015xua,Ackermann:2015zua,Li:2015kag}.

New physics beyond the standard model (SM) is needed to yield the main product $\tau \bar{\tau}$ in SM sector in WIMP annihilations. The leptophilic WIMPs were discussed in the literature \cite{Baltz:2002we,Chen:2008dh,Pospelov:2008jd,Cholis:2008qq,Fox:2008kb,Cao:2009yy,Bi:2009uj,Ibarra:2009bm,Kopp:2009et,Cohen:2009fz}. Here we consider that a new scalar mediates the interactions between the SM charged leptons and scalar/vectorial WIMPs (the annihilation of fermionic WIMPs is p-wave suppressed today), and the new couplings of the mediator to leptons are proportional to the lepton masses. If the scalar mediator is lighter than the WIMP mass, the way of a WIMP pair annihilating into an on-shell mediator pair is allowed (see e.g. Refs. \cite{Martin:2014sxa,Abdullah:2014lla,Rajaraman:2015xka,Cline:2015qha} for more). In this case, the scalar mediator's couplings to SM particles can be almost arbitrarily small.\footnote{There is a lower bound about the couplings, which is from the thermal equilibrium in the early universe.}  To fit the GeV gamma-ray excess and meanwhile evade present constraints from DM direct detections and collider experiments, we focus on the case that the mediator is lighter than the WIMP mass and the mediator-tau lepton coupling is much smaller than the mediator-WIMP coupling. Thus, the dominant annihilation mechanism of WIMPs is that a WIMP pair annihilates into an on-shell mediator pair which mainly decays to the heaviest leptons $\tau \bar{\tau}$. The case of $\tau \bar{\tau}$ mode dominant is naturally compatible with the antiproton spectrum observations from PAMELA \cite{Adriani:2010rc}, and is tolerant by the smooth positron spectrum of AMS-02 \cite{Bergstrom:2013jra,Hooper:2012gq,Ibarra:2013zia}.

A small scalar mediator-Higgs field mixing is discussed, and the mixing is small enough to keep $\tau \bar{\tau}$ dominant in the scalar mediator decays. With a small mixing introduced, one prospect is that the WIMP-target nucleus SI scattering may be detectable at the upgraded DM direct detection experiment in the next few years, and another prospect is that the scalar mediator may be observable in the future high-luminosity $e^+ e^-$ experiment. In fact, the small mediator-Higgs mixing can play an important role to the thermal equilibrium between DM and SM sectors in the early universe. The reaction rates of SM particles $\rightarrow$ WIMPs should be larger than the expansion rate of the universe for some time in the early universe, and this sets a lower bound about the couplings of the scalar mediator to SM particles. The lower bound of the coupling gives a lower limit on the cross section of WIMP-target nucleus SI scattering. These will be explored in this paper.

This article is organized as follows. After this introduction, the form of the interactions in new sector and the annihilation cross section of scalar/vectorial WIMPs are given in section II.
Next we give a detailed analysis about scalar WIMPs in section III, including the constraints and the test at future experiment. In section IV, we give a brief discussion about the test of vectorial WIMPs. The conclusions and some discussions are given in the last section.

\section{Interactions between WIMPs and SM}

In this article, we focus on scalar/vectorial WIMPs, with a new scalar field mediating the interactions between WIMPs and SM particles.

\subsection{The new sector interactions}

Consider that a real scalar field $\Phi$ mediates the interactions between scalar/vectorial WIMPs and SM particles, with $\Phi$ favoring SM leptons. Let us formulate the corresponding interactions. Following the forms in Refs \cite{Pospelov:2007mp,Batell:2012mj,Liu:2014cma,Abdallah:2015ter}, the effective interactions of $\Phi$ to scalar/vectorial WIMPs, charged lepton $l$ ($e$, $\mu$, $\tau$), Higgs field $H$ and $\Phi$ self-interactions are taken as
\begin{eqnarray}
\mathcal {L}_{S}^{\,i} &=& - \frac{\lambda}{2} \Phi^2 S^* S  - \mu \Phi S^* S  - \frac{\mu_3^{}}{3!} \Phi^3  - \frac{\lambda_4}{4!} \Phi^4  - \lambda_l^{} \Phi \bar{l} l \nonumber  \\
&& - \lambda' S^* S ( H^\dag H - \frac{v^2}{2} ) - \lambda_h \Phi^2 ( H^\dag H - \frac{v^2}{2} ) - \mu_h^{} \Phi ( H^\dag H - \frac{v^2}{2} )   \,, \label{scalar-DM}
\end{eqnarray}
\begin{eqnarray}
\mathcal {L}_{V}^{\,i} &=&  \frac{\lambda}{2} \Phi^2 V_\mu^* V^\mu  + \mu \Phi V_\mu^* V^\mu  - \frac{\mu_3^{}}{3!} \Phi^3  - \frac{\lambda_4}{4!} \Phi^4 - \lambda_l^{} \Phi \bar{l} l \nonumber  \\
&&  + \lambda' V_\mu^* V^\mu ( H^\dag H - \frac{v^2}{2} ) - \lambda_h \Phi^2 ( H^\dag H - \frac{v^2}{2} ) - \mu_h^{} \Phi ( H^\dag H - \frac{v^2}{2} )   \,, \label{vector-DM}
\end{eqnarray}
where $S$ is a scalar WIMP field, $V^\mu$ is a vectorial WIMP field, and a $Z_2$ symmetry is introduced to let WIMPs stable. The Yukawa type coupling parameter $ \lambda_l^{}$ is proportional to the charged lepton mass. $v$ is the vacuum expectation value with $v \approx$ 246 GeV, and $\Phi$ is chosen for no vacuum expectation value obtained \cite{Pospelov:2007mp,Batell:2012mj}. The parameter $\mu$ can be rewritten as $\mu = k m_S$, $\mu = k m_V$, with $k$ a dimensionless parameter and $m_S$, $m_V$ the scalar, vectorial WIMP mass respectively. The self-interaction terms of $\Phi$ are included, and the contribution from the cubic term $\Phi^3$ may need be considered in WIMP annihilations in some cases. The $\lambda'$ term is the DM-Higgs field interaction, which is also included for completeness.

The scalar component $h'$ of Higgs field and the scalar field $\Phi$ can mix after the electroweak symmetry breaking, giving the mass eigenstates $h$, $\phi$ in the form
\begin{eqnarray} \label{gauge1}
\left (
\begin{array}{c}
  h \\
  \phi
\end{array}
\right )=\left [
\begin{array}{cc}
           \cos\theta & \sin\theta  \\
           -\sin\theta & \cos\theta
         \end{array} \right ]
         \left (
\begin{array}{c}
  h' \\
  \Phi
\end{array}
\right ).
\end{eqnarray}
Here $\theta$ is the mixing angle, and one has
\begin{eqnarray}
\tan 2\theta =  \frac{2v \mu_h^{}}{m_{h'}^2 - m_\Phi^2}     \,.
\end{eqnarray}
For the Higgs sector being affected by $\Phi - H$ interactions as small as possible, here we suppose that the $\Phi - H$ interactions are relatively small, i.e. in the case of $ \lambda_h  \ll 1$ and $|v \mu_h^{} | \ll \min (m_\Phi^2, m_{h'}^2)$. For the mass eigenstates, one then has $m_\phi \simeq m_\Phi^{}$, $m_{h}^{} \simeq  m_{h'}^{}$. Thus, the value of $\theta$ can be very small, i.e. $| \sin\theta |$ $\sim |\theta| \ll 1$, $\cos\theta \sim 1$, and this is necessary to be compatible with experimental constraints. The $\theta$ value should be small enough to keep $\tau \bar{\tau}$ dominant in $\phi$'s decay, and this is essential to explain the Galactic center gamma ray excess.

Here we give a brief discussion about the $\lambda'$ term in Eqs. (\ref{scalar-DM}), (\ref{vector-DM}). In the case of $\lambda'$ being not much smaller than $\lambda$, $| \mu | / v$, the SM-like Higgs boson $h$ may have an appreciable contribution to the WIMP annihilations. Due to the definite Higgs-nucleon coupling, $h$ would have a significant contribution to the WIMP-nucleus scattering, while the Higgs portal DM is rigorously constrained by the direct detections. The circumstance may also occur that the annihilation of WIMPs is mainly via the interactions mediated by $\phi$, while the WIMP-nucleus scattering is mainly via the interactions mediated by $h$. As we focus on the new scalar $\phi$ portal DM, i.e. the case of $\lambda' \ll \lambda$, $\lambda' \ll | \mu | / v$, and a further request of $\lambda' \ll | \mu \sin\theta | / v$ to let the WIMP-nucleus scattering dominantly mediated by $\phi$. The contribution from $\lambda'$ term is neglected in this paper.

\subsection{Annihilations of WIMPs}

The case the scalar mediator is lighter than the WIMP mass $m_\phi < m_S$, $m_V$, is of our concern in this paper. When the coupling $\lambda$, $k^2 \gg |k \lambda_\tau|$, a WIMP pair predominantly annihilates into an on-shell $\phi \phi$ pair. The $\phi$ particle mainly decays to $\tau \bar{\tau}$, and the gamma rays from $\tau \bar{\tau}$ mode can reveal some properties of WIMPs. The differential gamma-ray flux from DM annihilation is
\begin{eqnarray}
E_\gamma^2 \frac{d \Phi_\gamma}{d E_\gamma} =  \frac{\langle\sigma_{ann} v_r \rangle_0  J }{8 \pi  m_{DM}^2} \sum_i BR_i E_\gamma^2 \frac{d N_\gamma^i}{d E_\gamma}    \, ,
\end{eqnarray}
where $\langle\sigma_{ann} v_r \rangle_0$ is the thermally averaged DM annihilation cross section today,
and $J$ is the annihilation $J-$factor. To fit the galactic center gamma-ray excess via $\tau \bar{\tau}$ mode as mentioned by the introduction (a WIMP pair annihilates into a $\tau \bar{\tau}$ pair, with WIMP mass about 7 -11 GeV and the annihilation cross section $\sim$  $0.5 \times 10^{-26}$ cm$^3/$s ), an alternative scheme is via the process of a WIMP pair $\rightarrow \phi \phi$ $\rightarrow (\tau \bar{\tau}) (\tau \bar{\tau})$, with the WIMP mass being twice $\sim 14-22$ GeV and the $\phi$ mass $m_\phi$ close to the WIMP mass $m_S, m_V$. In this case, the thermally averaged annihilation cross section today is $\sim 1 \times 10^{-26}$ cm$^3/$s (nearly $1 / 2$ of that at the thermally freeze-out temperature). The scheme above is of our concern.

\subsubsection{Scalar WIMPs}

\begin{figure}[!htbp]
\includegraphics[width=1.2in]{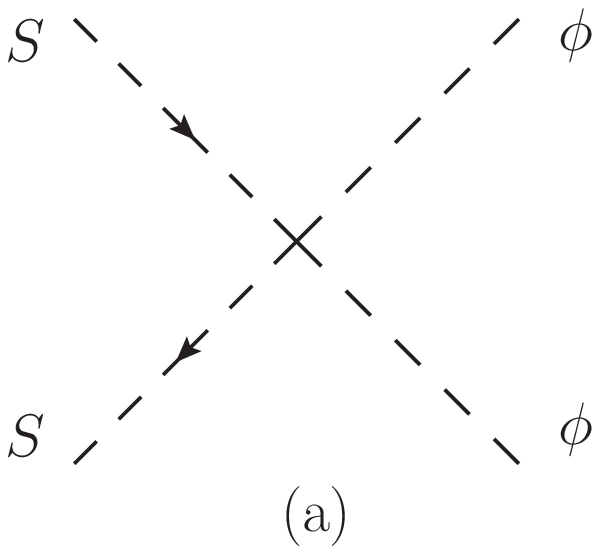} \vspace*{-1ex} \quad
\includegraphics[width=1.2in]{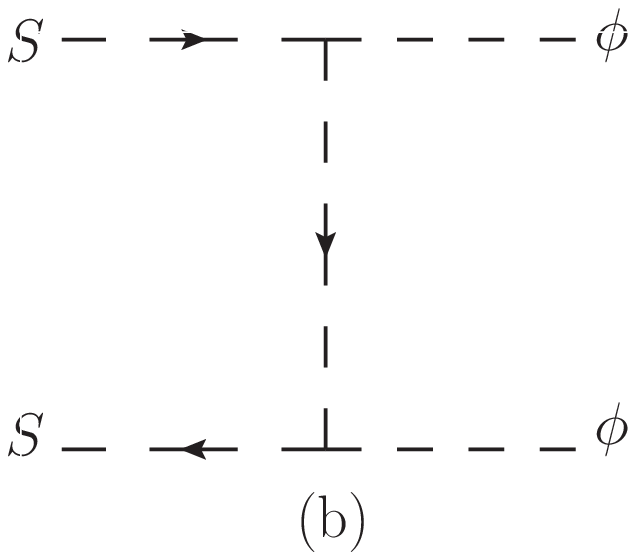} \vspace*{-1ex} \quad
\includegraphics[width=1.2in]{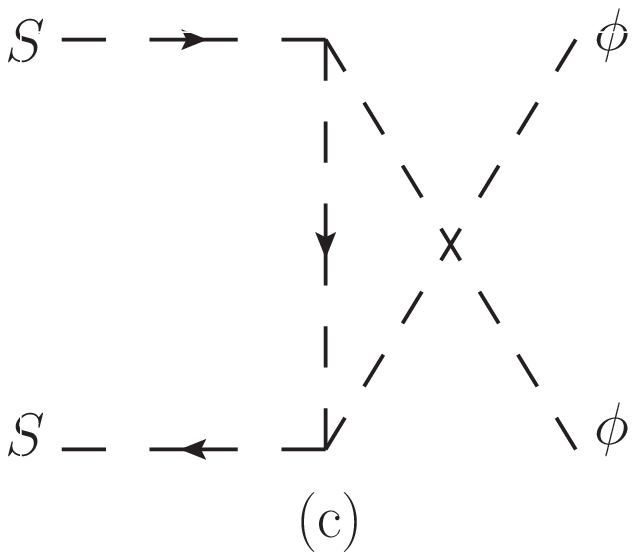} \vspace*{-1ex} \quad
\includegraphics[width=1.2in]{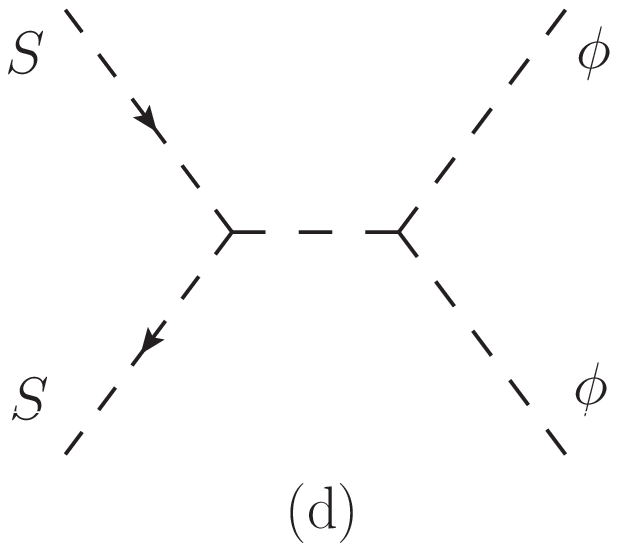} \vspace*{-1ex}
\caption{ The process of $S  S^*  \rightarrow \phi \phi$.}\label{s-phi}
\end{figure}

Let us consider the scalar WIMPs first. The process $S  S^*  \rightarrow \phi \phi$ is dominant in the WIMP annihilation, as shown in Fig. \ref{s-phi}. The WIMP annihilation cross section in one particle rest frame is
\begin{eqnarray}
\sigma_{ann} v_r \simeq  \frac{1}{2} \frac{\beta_f}{32 \pi (s - 2 m_S^2)}  (\lambda + 2 k^2 \frac{ m_S^2}{m_\phi^2 - 2 m_S^2 } + \frac{k k_3 m_S^{}  m_\phi^{}}{4 m_S^2 - m_\phi^2} )^2   \,. \label{sigvs}
\end{eqnarray}
Here the factor $\frac{1}{2}$ arises from the required $S  S^*$ type in annihilations, $v_r$ is the relative velocity of two WIMP particles, and $s$ is the total invariant mass squared. The parameter $\mu_3^{}$ in Eq. (\ref{scalar-DM}) is rewritten as $\mu_3^{} = k_3 m_\phi^{}$. $\beta_f$ is a kinematic factor, with
\begin{eqnarray}
\beta_f=\sqrt{1-\frac{4m_{\phi}^2}{s}} \, .
\end{eqnarray}
Due to the $\beta_f$ factor, when the mediator mass $m_\phi$ is slightly below the WIMP mass $m_S$, i.e. being close to the threshold of $S  S^*  \rightarrow \phi \phi$, the thermally averaged annihilation cross section $\langle\sigma_{ann} v_r \rangle_{0}^{}$ today (in the $T=0$ limit) is more suppressed in phase space compared with the cross section $\langle\sigma_{ann} v_r \rangle_{f}^{}$ at the freeze-out temperature $T_f^{}$. For this thermally freeze-out case, the key factor $\beta_f$ is failed to be expanded in Taylor series of $v_r^2$.

The present DM relic density $\Omega_D^{}$ and the parameter $x_f^{}$ (with $x_f^{} =$ $m_S^{}/T_f^{}$) can be approximately written as \cite{Kolb:1990vq, Griest:1990kh}
\begin{eqnarray}
\Omega_{D}^{} h^2 \simeq \frac{ 1.07 \times 10^9 \, \rm{GeV}^{-1}}{J_{ann} \sqrt{ g_\ast } m_{\rm Pl}^{} } \, ,
\end{eqnarray}
\begin{eqnarray}
x_f^{} \simeq \ln 0.038 \, c (c+2)  \frac{ g \, m_{\rm Pl}^{} m_{S}^{} \langle\sigma_{ann} v_r \rangle_{f}^{} }{ \sqrt{g_\ast  x_f^{}}}  \, , \label{xxf}
\end{eqnarray}
with
\begin{eqnarray}
J_{ann} =  \int_{x_{f}^{}}^{\infty}  \frac{\langle\sigma_{ann} v_r \rangle}{x^2} \rm{d}  \emph{x} \, .
\end{eqnarray}
Here $h$ is the Hubble constant (in units of 100 km/($ s \cdot $Mpc)), and $g_{*}$ is the number of the relativistic degrees of freedom with masses less than the temperature $T_f^{}$. $m_{\rm Pl}^{} $ is the Planck mass with the value $1.22 \times 10^{19} $ GeV, and $g$ is the degrees of freedom of DM. The parameter $c$ is of order one, and $c = 1/2$ is taken here. The thermally averaged annihilation cross section is \cite{Gondolo:1990dk,Cannoni:2013bza}
\begin{eqnarray}
\langle \sigma_{ann} v_r \rangle &=& \frac{2 x}{K_2^2 (x)}  \int_{0}^{\infty} \rm{d} \varepsilon \sqrt{\varepsilon} (1+2 \varepsilon) \nonumber \\
&&\times K_1(2 x \sqrt{1 + \varepsilon}) \sigma_{ann} v_r \,\, ,
\end{eqnarray}
with $\varepsilon =(s- 4 m_{S}^2)/4 m_{S}^2$. $K_i$ is the $i-$th order modified Bessel function.

\subsubsection{Vectorial WIMPs}

Now let us turn to the vectorial WIMPs. The process $V  V^*  \rightarrow \phi \phi$ is dominant in the vectorial WIMP annihilation. When $m_\phi$ is slightly below $m_V$, the annihilation cross section is
\begin{eqnarray}
\sigma_{ann} v_r \simeq  \frac{1}{2} \frac{\beta_f}{96 \pi (s - 2 m_V^2)}  (\lambda + 2 k^2 \frac{ m_V^2}{m_\phi^2 - 2 m_V^2 } +  \frac{k k_3 m_V^{}  m_\phi^{}}{4 m_V^2 - m_\phi^2})^2   \,. \label{sigvv}
\end{eqnarray}
For vectorial WIMPs, the thermally averaged annihilation cross section, the relic density are similar to the scalar case, with the corresponding parameter inputs in calculations.

\section{Analysis of Scalar WIMPs}

Here we give a detailed analysis about the scalar WIMPs, and the case of vectorial WIMPs is similar.

\subsection{The constraints of WIMP annihilations}

The process of WIMP pair $\rightarrow \phi \phi$ is dominant in WIMP annihilations. For the $\tau \bar{\tau}$ mode in WIMP annihilations, the present thermally averaged cross section set by the Galactic center gamma-ray excess is $\langle\sigma_{ann} v_r \rangle_0 \sim$  $1 \times 10^{-26}$ cm$^3/$s, with the mass of WIMPs in the range 14-22 GeV. The cold DM relic density today is $\Omega_{c}^{} h^2 =$ $0.1197 \pm 0.0022$ \cite{Ade:2015xua}. These constraints are taken to restrict the parameter spaces.

\begin{figure}[!htbp]
\includegraphics[width=3.2in]{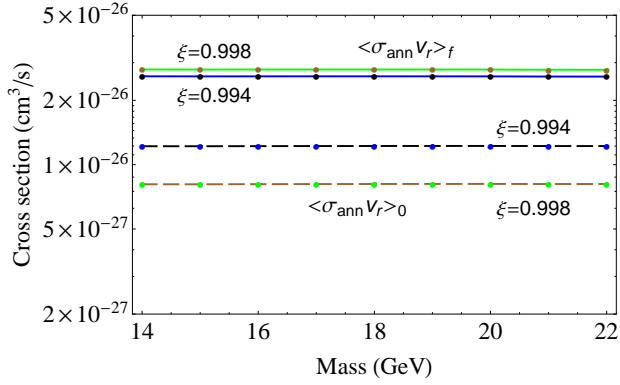} \vspace*{-1ex}
\caption{The thermally averaged annihilation cross section of WIMPs with masses in the range 14 - 22 GeV. The solid-dotted curves are the results of $\langle\sigma_{ann} v_r \rangle_{f}^{}$, with the lower one, the upper one corresponding to the case of $\xi =$0.994, $\xi =$0.998, respectively. The dashed-dotted curves are the results of $\langle\sigma_{ann} v_r \rangle_{0}^{}$, with the upper one, the lower one for the case of $\xi =$0.994, $\xi =$0.998, respectively.} \label{ann-cs}
\end{figure}

Define $m_\phi / m_S = \xi$, with $\xi < 1$ and $\xi$ close to 1. The factor $\beta_f$ plays a key role in fixing the ratio $\langle\sigma_{ann} v_r \rangle_{0}^{}$/$\langle\sigma_{ann} v_r \rangle_{f}^{}$, and the value of $\xi$ can be set by $\Omega_{c}^{} h^2$ and $\langle\sigma_{ann} v_r \rangle_0$. The numerical results of the thermally averaged annihilation cross sections are shown in Fig. \ref{ann-cs}, for WIMP masses in the range 14-22 GeV. When $\xi$ changes in the range $0.994 \lesssim$ $ \xi \lesssim 0.998$, $\langle\sigma_{ann} v_r \rangle_{0}^{}$ approximately varies from $1.2 \times 10^{-26}$ cm$^3/$s to $0.8 \times 10^{-26}$ cm$^3/$s. The derived annihilation cross section range of $\langle\sigma_{ann} v_r \rangle_{0}^{}$ together with the WIMP mass range of concern can give an interpretation about the Galactic center GeV gamma-ray excess.

Since the value of $\xi$ is obtained by the constraints, the coupling between WIMPs and the mediator $\phi$ is also determined. Taking $\xi \sim 0.994$, we have
\begin{eqnarray}
|\lambda + k k_3 / 3 - 2 k^2| \sim  2.0 \times 10^{-3} m_S  \rm{(GeV)}  \, ,
\end{eqnarray}
with $m_S$ in units of GeV. The coupling $\mu$ in the WIMP-$\phi$ trilinear term plays an important role in the WIMP-target nucleus scattering in DM direct detections. In the case of the WIMP-$\phi$ trilinear term dominating the WIMP annihilations, we have
\begin{eqnarray}
|k| = |\frac{\mu}{ m_S}| \sim  3.16 \times 10^{-2} \sqrt{m_S  \rm{(GeV)} }  \, .
\end{eqnarray}
If the contribution from $\lambda$, $\mu_3^{}$ ($= k_3 m_\phi^{}$) terms are significant in WIMP annihilations, the relation
\begin{eqnarray}
\lambda \sim  2 k^2 - k k_3 / 3 \pm  2.0 \times 10^{-3} m_S   \,  \label{value-l}
\end{eqnarray}
needs to be taken care of. In the case of a large $\lambda$ value together with a comparable large $|k|$ and a mall $|k_3|$ value, or a large $|k_3|$ value together with a comparable large $|k|$ and a mall $\lambda$ value, the s-channel annihilation of WIMP directly annihilating into SM particles can be enhanced. As we focus on the case of such direct annihilation being suppressed in this paper, e.g. the value of $\lambda$ $+ |k k_3| / 3 $ $+ 2 k^2$ being order of $2.0 \times 10^{-3} m_S$, and here a range of $k$
\begin{eqnarray}
k^2  \lesssim  1.4 \times 10^{-3} m_S \rm{(GeV)}  \, ,  \label{value-k}
\end{eqnarray}
is considered in calculations.

\subsection{The constraints of $\phi$}

Now we give a brief discussion about the coupling of $\phi$ to SM particles, that is, the $\lambda_\tau$'s value and the mixing angle $\theta$. Some parameters are inputted as follows,  $m_{\tau}=1.77682$ GeV, $ m_{\mu} = 0.105658 $ GeV, $m_t = 173.21 $ GeV, $m_{b}= 4.18$ GeV, with the results from PDG \cite{Agashe:2014kda}.

\subsubsection{The $\lambda_\tau$ value}

As discussed above, the case of s-channel suppressed in WIMP annihilations is of our concern, i.e. $\lambda, k^2$ $\gg |k \lambda_\tau|$. Here the $k$ value in the case of the $\mu$ term dominant is taken to restrict the $\lambda_\tau$'s value, that is
\begin{eqnarray}
\lambda_\tau \ll 3.16 \times 10^{-2} \sqrt{m_S  \rm{(GeV)} }  \, . \label{g-c}
\end{eqnarray}
The decay width of $\phi$ is
\begin{eqnarray}
\Gamma_\phi \simeq  \frac{m_\phi}{8 \pi} [ \lambda_\tau^2 ( 1-\frac{4m_{\tau}^2}{m_{\phi}^2} )^{3/2}  + 3 \frac{ m_b^2}{v^2} \sin^2\theta  ( 1-\frac{4m_{b}^2}{m_{\phi}^2} )^{3/2} ] \, , \label{phi-decay}
\end{eqnarray}
with $\lambda_\tau$ term dominant. Taking the limit of $\lambda_\tau$ in Eq. (\ref{g-c}), we have $\Gamma_\phi$ $ \ll (m_S - m_\phi)$. Thus, the $\xi$'s range of concern is feasible.

The $\phi$ particle contributes to the muon $g-2$, and the one-loop result is \cite{Hektor:2015zba}
\begin{eqnarray}
a_\mu^\phi  \simeq  \frac{\lambda_\mu^2}{8 \pi^2}\frac{m_\mu^2 }{m_\phi^2 } (\ln \frac{m_\phi^2 }{m_\mu^2 } -\frac{7}{6})\, . \label{g-2}
\end{eqnarray}
The difference between experiment and theory is \cite{Agashe:2014kda}
\begin{eqnarray}
\Delta a_\mu =  a_\mu^{exp} -  a_\mu^{SM} = 288 (63)(49) \times 10^{-11}\, . \label{g-result}
\end{eqnarray}
Taking the replacement $\lambda_\mu = \lambda_\tau m_\mu / m_\tau$ and $m_\phi \sim m_S$ in Eq. (\ref{g-2}), we can find that the upper limit of $\lambda_\tau$ in Eq. (\ref{g-c}) is tolerant by the muon $g-2$ result.

\subsubsection{The mixing angle $\theta$}

For the WIMP mass range of concern, according to Eq. (\ref{phi-decay}), if the $b \bar{b}$ channel is not larger than 20\%, the $\theta$ value should satisfy the relation $|\sin\theta| \lesssim 20 \lambda_\tau$ (when $\lambda_\tau$ is compatible with 0.01, this constraint is relaxed). The Higgs hunt results at LEP \cite{Barate:2003sz} set an upper limit on $\theta$,
\begin{eqnarray}
\sin^2\theta \lesssim 0.1 ~ (\phi \rightarrow \tau \bar{\tau}), \quad \sin^2\theta \mathcal {B}^{}_{\phi \rightarrow b \bar{b}} \lesssim 2 \times 10^{-2} ~ .
\end{eqnarray}
We can see that the constraints from LEP are mild. The ATLAS \cite{ATLAS:2011cea} and CMS \cite{Chatrchyan:2012am} search constraints about light Higgs-like particles can be approximately written as \cite{Clarke:2013aya,Haisch:2016hzu}
\begin{eqnarray}
\sin^2\theta \mathcal {B}^{}_{\phi \rightarrow \mu^+ \mu^-} \lesssim  \mathcal {B}^{SM}_{h \rightarrow \mu^+ \mu^-} ~ .
\end{eqnarray}
With these constraints, we obtain an upper limit of $\theta$
\begin{eqnarray}
\sin^2\theta  \lesssim  6 \times 10^{-2} ~, \quad \rm{and} \quad  |\sin\theta| \lesssim 20 \lambda_\tau \, .
\end{eqnarray}
The constrains of the DM direct detection and Higgs boson decay will be discussed in the following.

\subsection{Thermal equilibrium constraints}

In the early universe, the WIMPs and SM particles are in thermal equilibrium. The reaction rates of WIMP pairs $\leftrightarrow$ SM particles exceed the expansion rate of the universe for some time,
\begin{eqnarray}
\langle \sigma_{ann} v_r \rangle n_{eq} \gtrsim 1.66 \frac{\sqrt{g_\ast} ~ T^2}{m_{\rm Pl}^{}} \label{rear} \,\, ,
\end{eqnarray}
where $n_{eq}$ is the corresponding equilibrium number density, with $n_{f} = 3 \zeta (3) g_f T^3 / 4 \pi^2$ for fermions in the relativistic limit. For SM particles $\rightarrow$ WIMP pairs, the annihilation cross section of each fermion specie is
\begin{eqnarray}
 \sigma_{ann} v_r = \frac{\lambda_{SM}^2 k^2 \sqrt{1- 4 m_S^2 / s}}{32 \pi (s - 2 m_f^2)} \frac{m_S^2 (s - 4 m_f^2)}{(s -   m_\phi^2)^2}  \,\, ,
\end{eqnarray}
with $\lambda_{SM} \simeq \lambda_{l}, \sin\theta m_q / v$ for charged leptons, quarks, respectively. The reaction rate can set a lower bound about the $\phi$'s coupling to SM particles (see e.g. Refs. \cite{Chu:2011be,Dolan:2014ska} for more). If the mixing angle  $\theta$ is tiny, with the contribution mainly from the $\tau \bar{\tau}$ annihilation, the reaction rate can give a lower bound on $\lambda_\tau$. However, in this case, the WIMPs are insensitive in target nucleus scattering detections, and traces of $\phi$ are difficult to be observed at collider experiment. Here we focus on the interesting case of $t \bar{t}$ contribution dominating the SM particle reaction rate at $T \sim m_t$. By the calculation, we can obtain that the $t \bar{t}$ contribution is dominant when $|\sin\theta|$ is some times larger than $\sqrt{10} \lambda_\tau$ (this value corresponding to the nearly equal contributions of $t \bar{t}$ and $\tau \bar{\tau}$), e.g. $|\sin\theta| \gtrsim$ $10 \lambda_\tau$. Moreover, an appreciable $\sin\theta$ value is available in interpreting the Galactic center gamma ray excess. Now, we have a range of $\theta$,
\begin{eqnarray}
10 \lambda_\tau \lesssim |\sin\theta| \lesssim 20 \lambda_\tau   \, .  \label{theta-range}
\end{eqnarray}

For the case of the $t \bar{t}$ contribution dominating the SM reaction rate, according to Eq. (\ref{rear}), we obtain that the constraint can be written as
\begin{eqnarray}
\sin^2\theta  k^2 m_S^2  ({GeV}) \gtrsim 2.2 \times 10^{8} \frac{\pi^3 \sqrt{g_\ast} ~ m_t}{\zeta (3) m_{\rm Pl}^{}} \approx 8.5 \times 10^{-7} \,\, . \label{t-e-top}
\end{eqnarray}
This constraint is taken as the lower bound for the mixing angle $\theta$ and the parameter $k$. In fact, the constraint is valid for the WIMPs in a general mass range of $m_\phi < m_S \ll m_t$.

\subsection{DM direct detection}

\begin{figure}[!htbp]
\includegraphics[width=3.2in]{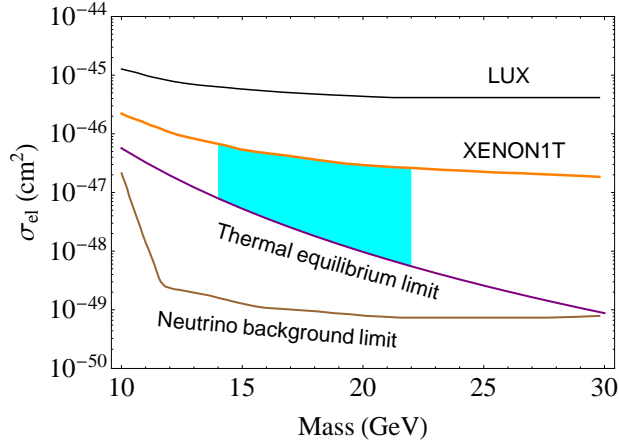} \vspace*{-1ex}
\caption{The SI elastic cross section $\sigma_{\rm {el}}$ of WIMPs in a mass range 10 - 30 GeV. The solid curves from top to bottom are the upper limit set by LUX \cite{Akerib:2015rjg}, the upper limit set by XENON1T \cite{Aprile:2015uzo}, the lower limit set by the thermal equilibrium, the lower detection limit set by the neutrino background \cite{Billard:2013qya}, respectively. The filled area is the allowed region of the elastic cross section in the potential mass range $m_S \sim$ 14 - 22 GeV of concern.} \label{dm-n-cs}
\end{figure}

Here we turn to the direct detection of WIMPs. The WIMP-target nucleus scattering is mainly mediated by $\phi$. The effective coupling between $\phi$ and nucleon can be set by $\sin\theta g_{hNN}^{}$, and $g_{hNN}^{}$ is the Higgs-nucleon coupling, with $g_{hNN}^{} \simeq 1.71 \times 10^{-3}$ \cite{He:2008qm} adopted here.\footnote{There is an uncertainty about the value of the Higgs-nucleon coupling. See e.g. Refs \cite{Ellis:2000ds,Gondolo:2004sc,Alarcon:2011zs,Cheng:2012qr} for more.}  The cross section of the WIMP-nucleon SI elastic scattering is
\begin{eqnarray}
\sigma_{\rm {el}} \simeq \frac{\sin^2\theta k^2 m_S^2 g_{hNN}^2 m_N^2}{4 \pi (m_S + m_N)^2 m_\phi^4} \,,
\end{eqnarray}
where $m_N^{}$ is the nucleon mass.

For WIMPs in the mass range of concern, the recent DM searching results of LUX \cite{Akerib:2015rjg} and XENON1T \cite{Aprile:2015uzo} set stringent upper limits on the mixing angle $\theta$ and the parameter $k$. In addition, the thermal equilibrium condition requirement of Eq. (\ref{t-e-top}) gives a lower bound on the parameters. Taking $m_\phi \sim m_S$, and considering the neutrino background \cite{Billard:2013qya} in detections, the tolerant hunting region of the cross section $\sigma_{\rm {el}}$ is depicted in Fig. \ref{dm-n-cs}, for WIMPs in an ordinary mass range 10 - 30 GeV. The filled region is for the potential mass range $m_S \sim$ 14 - 22 GeV of concern, which is indicated by the galactic center gamma ray excess, and the allowed region of the cross section is $\sigma_{\rm {el}} \sim$ $10^{-48} - 10^{-46}$ $\rm{cm}^2$. The parameter spaces are set by the thermal equilibrium limit and the recent XENON1T results. For $m_S \sim$ 14 - 22 GeV, the upper limit of XENON1T is fitted in the form
\begin{eqnarray}
a \times (m_S)^b({GeV}) \times \frac{ g_{hNN}^2 m_N^2}{4 \pi (m_S + m_N)^2 m_S^4} \,,
\end{eqnarray}
with the fitting values $a = (3.89 \pm 0.98 ) \times 10^{-10}$  , $b = 3.71 \pm 0.09$. Thus, in the WIMP mass range of concern, the constraints of $\theta$ and $k$ can be expressed as
\begin{eqnarray}
8.5 \times 10^{-7}  \lesssim \sin^2\theta  k^2 m_S^2  ({GeV}) \lesssim  2.61 \times 10^{-5} (\frac{m_S }{20})^{3.71}  \,\, . \label{thermal-Xenon}
\end{eqnarray}
This is the parameter space allowed, and it is detectable in DM direct detections in the future.

\subsection{New sector search at collider}

\subsubsection{New channels for Higgs decays}

After the discovery of the SM-like Higgs boson at LHC \cite{Aad:2012tfa,Chatrchyan:2012xdj}, the exploration of the Higgs portal new physics attracts much attention in recent years. In our scheme, the Higgs boson can decay into a WIMP pair, and the decay width is
\begin{eqnarray}
\Gamma_{h\rightarrow S^*  S} =  \frac{\sin^2\theta  k^2 m_S^2}{16 \pi m_h} \sqrt{1-\frac{4m_{S}^2}{m_h^2}} \, .
\end{eqnarray}
Taking $m_h =$ 125 GeV \cite{Aad:2015zhl}, the total width of SM Higgs is $\Gamma_{h \rm(SM)} =$ $4.07 \times 10^{-3}$ GeV \cite{Denner:2011mq,Agashe:2014kda}. With the constraints of the thermal equilibrium limit and the recent XENON1T results, i.e. Eq. (\ref{thermal-Xenon}), the branching ratio of Higgs boson decaying into a scalar WIMP pair is
\begin{eqnarray}
3.3 \times10^{-8}  \lesssim \mathcal {B}_{h\rightarrow S^*  S} \lesssim 1.0 \times 10^{-6} (\frac{m_S }{20})^{3.71} \,.
\end{eqnarray}
This invisible branching ratio is very small and difficult to investigate at present and in the future collider experiment.

According to Eq. (\ref{scalar-DM}), the channel of Higgs boson decaying into a on-shell $\phi \phi$ pair is allowed for the $\phi$ mass of concern. For $| \sin\theta |$ $\sim |\theta| \ll 1$, $\cos\theta \sim 1$, the decay width of $h \rightarrow \phi  \phi$ can be approximately written as
\begin{eqnarray}
\Gamma_{h \rightarrow \phi  \phi} \approx  \frac{\lambda_h^2 v^2 \cos^6\theta  }{32 \pi m_h^{}} \sqrt{1-\frac{4m_{\phi}^2}{m_h^2}} \, ,
\end{eqnarray}
with the $\sin\theta$ terms neglected. This channel should be small compared with the SM leading channel $h \rightarrow b  \bar{b}$, i.e. $\lambda_h v^2 \ll \sqrt{6} m_b m_h$. As a rough estimate, an upper limit $\lambda_h v^2 \lesssim  m_b m_h / 4$ is taken in discussions (i.e. the decay width $\Gamma_{h \rightarrow \phi  \phi} \lesssim 2.2 \times 10^{-5}$ GeV).

The decay channel $h \rightarrow \phi  \phi$ may be detectable at the future precise Higgs decay measurement via the process $h \rightarrow \phi  \phi$ $\rightarrow (\tau \bar{\tau}, b  \bar{b})$  $(\tau \bar{\tau}, b  \bar{b})$. According to Eqs. (\ref{phi-decay}), (\ref{theta-range}), the branching ratios of the two main decay channels of $\phi$ are
\begin{eqnarray}
6.4 ~ \%  \lesssim \mathcal {B}_{\phi \rightarrow  b  \bar{b}} \lesssim 21 ~ \% \,, \quad  \mathcal {B}_{\phi \rightarrow  \tau \bar{\tau}} \simeq 1 - \mathcal {B}_{\phi \rightarrow  b  \bar{b}} \, ,
\end{eqnarray}
with $m_{\phi}^{} = 20 $ GeV adopted as input. Thus, the upper limits of the exotic branching ratios in $h$ decay are as follows:
\begin{eqnarray}
&&\mathcal {B}_{h \rightarrow \phi  \phi \rightarrow (\tau \bar{\tau})(\tau \bar{\tau})} \simeq  \frac{\Gamma_{h \rightarrow \phi  \phi} ~ \mathcal {B}_{\phi \rightarrow  \tau \bar{\tau}}^{2} }{\Gamma_{h \rm{(SM)}} + \Gamma_{h \rightarrow \phi  \phi}} \lesssim (3.4 - 4.7) \times 10^{-3}  \, ,  \\
&&\mathcal {B}_{h \rightarrow \phi  \phi \rightarrow (\tau \bar{\tau})( b  \bar{b})} \simeq  \frac{2 \Gamma_{h \rightarrow \phi  \phi} ~ \mathcal {B}_{\phi \rightarrow  \tau \bar{\tau}}  \mathcal {B}_{\phi \rightarrow  b  \bar{b}}}{\Gamma_{h \rm{(SM)}} + \Gamma_{h \rightarrow \phi  \phi}} \lesssim (0.64 - 1.8) \times 10^{-3} \, ,  \\
&&\mathcal {B}_{h \rightarrow \phi  \phi \rightarrow ( b  \bar{b})( b  \bar{b})} \simeq  \frac{\Gamma_{h \rightarrow \phi  \phi} ~ \mathcal {B}_{\phi \rightarrow  b \bar{b}}^{2} }{\Gamma_{h \rm{(SM)}} + \Gamma_{h \rightarrow \phi  \phi}} \lesssim (0.22 - 2.4) \times 10^{-4} \, .
\end{eqnarray}
Due to the missing neutrino(s) in $\tau$ decay, the resolution of $m_{\tau^+ \tau^-}$ is poor (about 15\%) \cite{Agashe:2014kda}. Thus, the above three channels are comparable in the Higgs decay search. As the $e^+ e^-$ collider has a more clean environment compared with the hadron collider, here we focus on the precise tests of Higgs decay channels at the future $e^+ e^-$ collider. At the center of mass energy $\sqrt{s} \simeq$ 250 - 350 GeV, the dominant Higgs production mechanism is via the Higgs-strahlung process $e^+ e^- \rightarrow Z \phi $, and this can be employed for the precise measurement of the Higgs decays. The cross sections of Higgs-strahlung mode in $e^+ e^-$ collisions at $\sqrt{s} = 250, 350$ GeV are 211, 134 fb \cite{Dawson:2013bba}, respectively. For $\sqrt{s} = 250 $ GeV, there are about $ 10^5 -  10^6$ Higgs events produced at a high integrated luminosity of 500 fb$^{-1} $ $-$ 5 ab$^{-1}$. In this case, if the decay width of $h \rightarrow \phi  \phi$ is near the upper limit, there will be tens $-$ hundreds tagging events of $h \rightarrow \phi  \phi$ via the three decay modes discussed above. Thus, the $h \rightarrow \phi \phi$ decay can be investigated at the future Higgs factory, or the corresponding limit is set by the experiment.

\subsubsection{Production of $\phi$ at collider}

\begin{figure}[!htbp]
\includegraphics[width=3in]{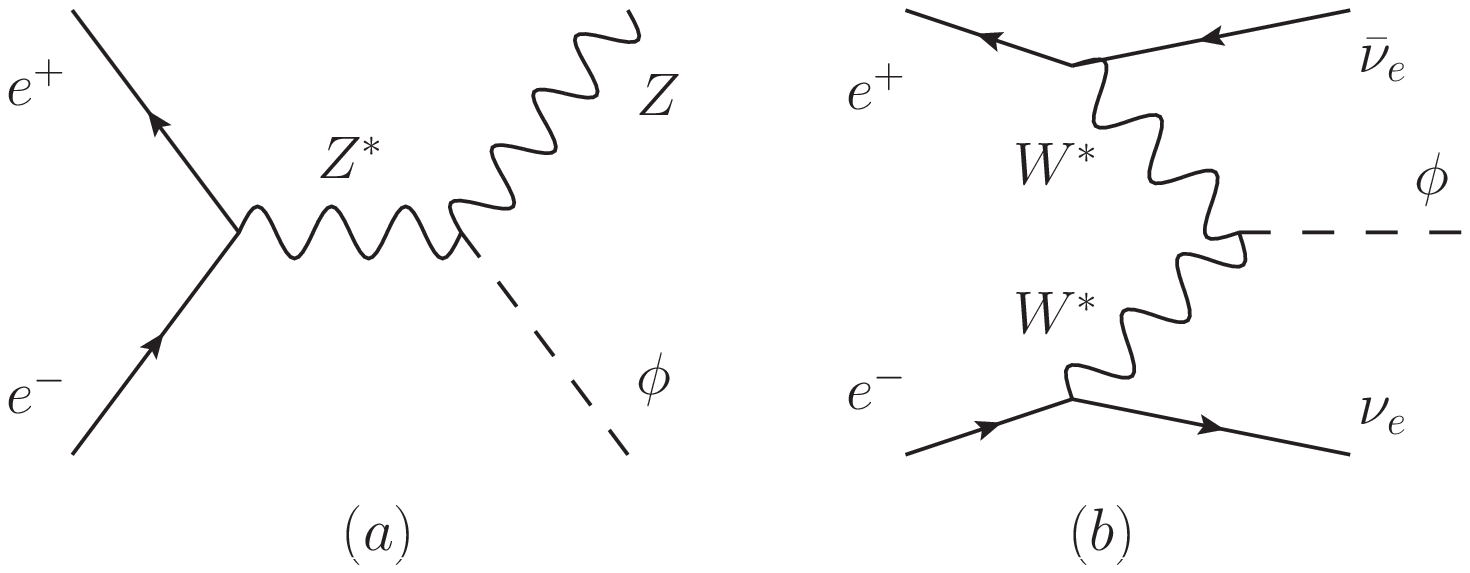} \vspace*{-1ex} \quad
\includegraphics[width=1.3in]{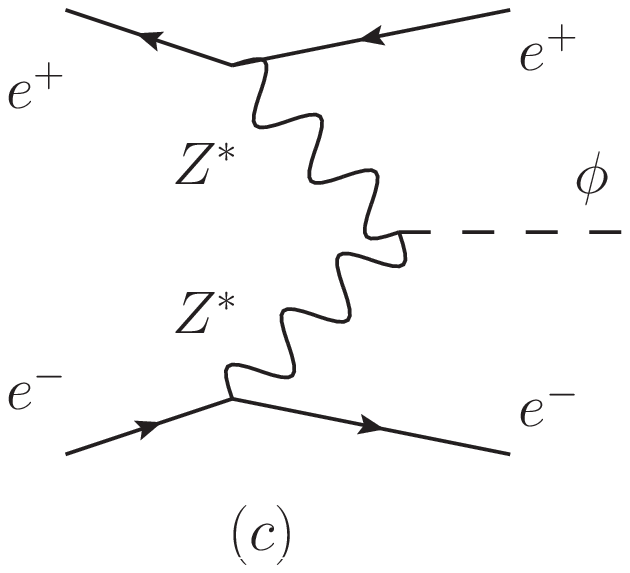} \vspace*{-1ex}
\caption{The main production processes of $\phi$ at $e^+ e^-$ collider.} \label{ph-e-e}
\end{figure}

Now we turn to the $\phi$ production at collider. Due to the messy background at the hadron collider, the constraints from ATLAS \cite{ATLAS:2011cea} and CMS \cite{Chatrchyan:2012am} are mild on the teens/tens GeV $\phi$ of concern, as discussed above. Here we focus on the search of $\phi$ at high energy $e^+ e^-$ collider, and this clean environment machine is good for high precise studies. The dominant production processes of $\phi$ are the $\phi -$strahlung, the $W W$ fusion, and the $Z Z$ fusion, as depicted in Fig. \ref{ph-e-e}.

The $\phi$ strahlung process is similar to the case of Higgs boson production, and the corresponding cross section can be written as
\begin{eqnarray}
\sigma_{e^+ e^- \rightarrow Z \phi} =  \frac{\sin^2\theta  G_F^2 m_Z^4 }{96 \pi s}  (v_e^2 + a_e^2) \beta \frac{\beta^2 + 12 m_Z^2/s}{(1 - m_Z^2/s)^2} \, ,
\end{eqnarray}
where $G_F^{} = 1.1663787(6) \times 10^{-5}$ $\rm{GeV}^{-2}$ \cite{Agashe:2014kda} is the Fermi coupling constant, and $v_e = -1 +4\sin^2\theta_W$, $a_e = -1$ are the vector, axial-vector current parameters, respectively. $\beta$ is the phase space factor, with
\begin{eqnarray}
\beta = \sqrt{(1-\frac{m_{\phi}^2}{s} -\frac{m_{Z}^2}{s})^2 - \frac{4m_\phi^2 m_Z^2}{s^2}} \, .
\end{eqnarray}
The cross section of the vector boson ($W W$, $Z Z$) fusion process can be written in the form \cite{Zerwas:desy,Kilian:1995tr,Djouadi:1996uj}
\begin{eqnarray}
&&\sigma = \frac{\sin^2\theta  G_F^3 M_v^4 }{64 \sqrt{2} \pi^3} \int^1_{\kappa_\phi} d x \int^1_{x} \frac{d y}{[1+(y-x)/{\kappa_v}]^2} [(\hat{v}^2 + \hat{a}^2)^2 f(x,y) + 4\hat{v}^2 \hat{a}^2 g(x,y)] \, ,\\
&&f(x,y)= ( \frac{2 x}{y^3} - \frac{1+2x}{y^2} +\frac{2+x}{2y} -\frac{1}{2})[\frac{z}{1+z} -\log(1+z)] + \frac{ x}{y^3}\frac{z^2 (1-y)}{1 + z}                    \nonumber   \\
&&g(x,y)= (  - \frac{ x}{y^2} +\frac{2+x}{2y} -\frac{1}{2})[\frac{z}{1+z} -\log(1+z)]         \nonumber
\end{eqnarray}
with $M_v^{} =$ $m_{W}^{},  m_{Z}^{}$ for the $W$, $Z$ boson respectively, $\kappa_\phi = m_{\phi}^2/s$, $\kappa_v = M_v^2/s$, and $z = y (x- \kappa_\phi)/(x \kappa_v)$. $\hat{v}$,  $\hat{a}$ are the electron couplings to the vector bosons, with $\hat{v} =$ $\hat{a} = \sqrt{2}$ for the $W$ boson, and $\hat{v} = v_e$, $\hat{a} =a_e$ for the $Z$ boson.

Let us give a further discussion about the range of $\theta$ before evaluating the production cross section of $\phi$.  From Eq. (\ref{value-k}) and Eq. (\ref{thermal-Xenon}), we can derive a lower bound $|\theta| \gtrsim |\theta_l|$, with
\begin{eqnarray}
\frac{6.1 \times 10^{-4}}{m_S^3} \lesssim \sin^2\theta_l \lesssim \frac{1.9 \times 10^{-2}}{m_S^3} (\frac{m_S }{20})^{3.71}   \, .
\end{eqnarray}
In addition, $|\theta|$ should be much smaller than 1. As a rough estimate, an alteration of order $0.1\%$   about the Higgs production and decay  is tolerant by the present experiment. Here, an upper limit $\sin^2\theta \lesssim 10^{-3}$ is taken. In SM particle scattering processes, the contribution from Higgs boson keeps dominant among the $\phi, h$'s contributions.

\begin{figure}[!htbp]
\includegraphics[width=3.2in]{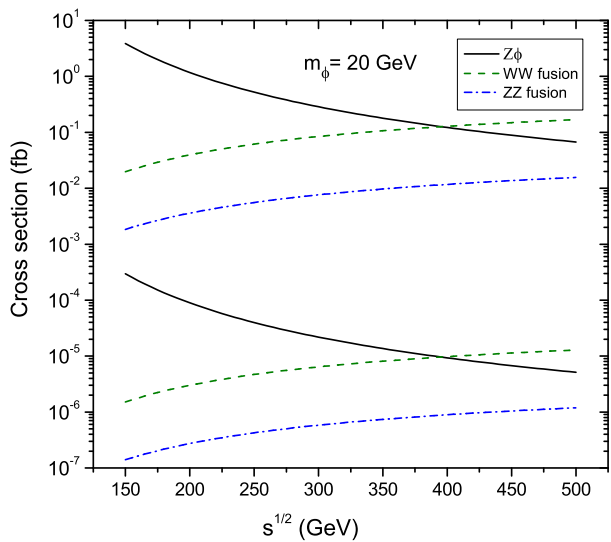} \vspace*{-1ex}
\caption{The production cross section of $\phi$ at $e^+ e^-$ collider with $m_\phi =$ 20 GeV and $\sqrt{s}$ varying in a range of 150 - 500 GeV. The solid curves, the dashed curves and the dashed dotted curves are the production cross sections of the $\phi$ strahlung, $W W$ fusion, and $Z Z$ fusion, respectively. In each type curve, the upper one, the lower one are for the case of $\sin^2\theta = 10^{-3}$, $\sin^2\theta = {6.1 \times 10^{-4}}/{m_S^3}$, respectively. } \label{ph-pro-s}
\end{figure}

\begin{figure}[!htbp]
\includegraphics[width=3.2in]{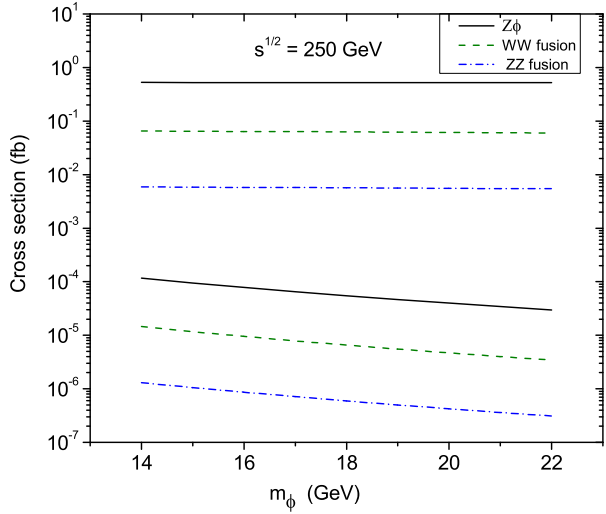} \vspace*{-1ex}
\caption{The $\phi$ production cross section as a function of $m_\phi$ with $m_\phi$ varying in a range of 14 - 22 GeV at $\sqrt{s} =$ 250 GeV. The solid curves, the dashed curves and the dashed dotted curves are the production cross sections of the $\phi$ strahlung, $W W$ fusion, and $Z Z$ fusion, respectively. In each type curve, the upper one, the lower one are for the case of $\sin^2\theta = 10^{-3}$, $\sin^2\theta = {6.1 \times 10^{-4}}/{m_S^3}$, respectively.} \label{ph-prod-m}
\end{figure}

With $m_\phi \sim m_S$ in this paper, we consider the production of $\phi$ in the range $m_\phi \sim$ 14 - 22 GeV. Fixing $m_\phi =$ 20 GeV, the dependence of the $\phi$ strahlung, $W W$ fusion, and $Z Z$ fusion cross sections with the center of mass energy $\sqrt{s}$ is depicted in Fig. (\ref{ph-pro-s}), for $\sqrt{s}$ varying in a range 150 - 500 GeV. The upper limit, lower limit of the production cross sections are corresponding to $\sin^2\theta = 10^{-3}$, $\sin^2\theta = {6.1 \times 10^{-4}}/{m_S^3}$, respectively. It can be seen that, for $\sqrt{s}$ below 400 GeV, the main production process of $\phi$ is the $\phi$ strahlung mechanism. At a given center of mass energy $\sqrt{s} =$ 250 GeV (the potential Higgs production energy), the production cross sections of $\phi$ for $\phi$ in the range 14 - 22 GeV is shown in Fig. (\ref{ph-prod-m}), with the same upper limit, lower limit in the processes of $\phi$ strahlung, $W W$ fusion, and $Z Z$ fusion as that of Fig. (\ref{ph-pro-s}). At the same $\sin^2\theta$ value, the cross section changes slowly with $m_\phi$. Thus, the upper limit results of $\phi$ production cross section given in Fig. (\ref{ph-pro-s}) are roughly the cross section of $\phi$ with the mass of 14 - 22 GeV.

\begin{figure}[!htbp]
\includegraphics[width=3in]{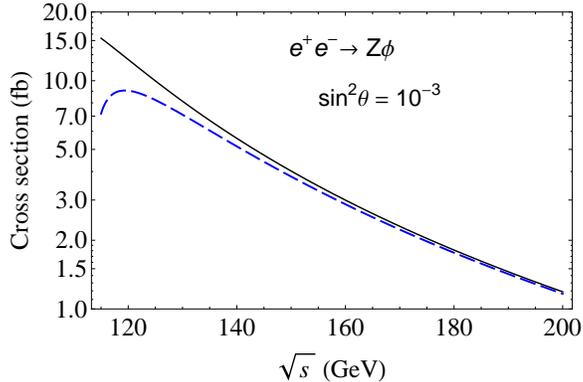} \vspace*{-1ex}
\caption{The production cross section of $\phi$ as a function of $\sqrt s$ in $\phi$ strahlung process at $e^+ e^-$ collider. The value $\sin^2\theta = 10^{-3}$ is taken here, and $\sqrt s$ varies in a range of 115 - 200 GeV. The solid curve, the dashed curve are for the case of $m_{\phi} =$ 14 GeV, $m_{\phi} =$ 22 GeV, respectively. } \label{ph-lowE}
\end{figure}

If the $|\theta|$ value is near the upper limit of the parameter space, the signature of $\phi$ may appear at the future high luminosity $e^+ e^-$ collider. Considering $\sqrt{s}$ below 400 GeV, the main production mechanism of $\phi$ is via the $\phi$ strahlung. In the case $\sin^2\theta = 10^{-3}$, there is about a hundred $\phi$ events produced at $\sqrt{s} =$ 250 GeV with a integrated luminosity of 200 fb$^{-1}$. The dominant final state of $\phi$ is $\tau \bar{\tau}$, and the second branching fraction of $b \bar{b}$ final state is about 6.4\% $-$ 21\%. The $\phi$ particle can be searched by the final states ($\phi\rightarrow$ $\tau \bar{\tau}$)($Z \rightarrow$ $q \bar{q}$), ($\phi\rightarrow$ $b \bar{b}$)($Z \rightarrow$ $l \bar{l}, q \bar{q}$). In fact, as shown in Fig. (\ref{ph-lowE}), it is better to test the non-standard model $\phi$-like particle at a low center of mass energy collider with a high luminosity, e.g. $\sqrt{s} \sim$ 120 - 150 GeV with the energy above the $Z \phi$ production threshold. For $\sin^2\theta = 10^{-3}$, there are about 800 ($\sqrt{s} =$150 GeV) - 2000 ($\sqrt{s} =$120 GeV) $\phi$ production events with a integrated luminosity of 200 fb$^{-1}$. Thus, the new particle $\phi$ with $|\theta|$ near the upper limit of the parameter space can leave traces at the future high luminosity $e^+ e^-$ collider, or the upper limit of $|\theta|$ is reduced by the search result.

\section{Analysis of vectorial WIMPs}

The vectorial WIMPs is similar to the case of scalar WIMPs. To satisfy the corresponding constraints, the value of $\xi$ is approximately in the same range as the scalar WIMPs. The cross section of the vectorial WIMP-nucleon SI elastic scattering is
\begin{eqnarray}
\sigma_{\rm {el}} \simeq \frac{\sin^2\theta k^2 m_V^2 g_{hNN}^2 m_N^2}{4 \pi (m_V + m_N)^2 m_\phi^4} \,.
\end{eqnarray}
In the following, we just focus on the significant differences for vectorial WIMPs, and give a brief discussion about them.

For vectorial WIMPs, with $\xi \sim$ 0.994, we have
\begin{eqnarray}
|\lambda + k k_3 / 3 - 2 k^2| \sim \sqrt{3} \times 2.0  \times 10^{-3} m_V^{}  \rm{(GeV)}  \, .
\end{eqnarray}
In the thermal equilibrium era of the early universe, the reaction rates of SM particles $\rightarrow$ WIMP pairs exceed the expansion rate of the universe. The cross section of each SM fermion specie annihilating into a vectorial WIMP pair is
\begin{eqnarray}
 \sigma_{ann} v_r = \frac{\lambda_{SM}^2 k^2 \sqrt{1- 4 m_V^2 / s}}{32 \pi (s - 2 m_f^2)} \frac{m_V^2 (s - 4 m_f^2)}{(s -   m_\phi^2)^2} [ 2 + \frac{(s - 2 m_V^2)^2}{4 m_V^4} ]  \,\, .
\end{eqnarray}
At $T \sim m_t$, the reaction rates of SM particles are significant enhanced by the longitudinal polarization of vectorial WIMPs, e.g. the enhancement over $10^{5}$ for $t \bar{t}$. Consider the $t \bar{t}$ contribution dominating the SM particle reaction rate, and we have a range of $\theta$,
\begin{eqnarray}
2 \lambda_\tau \lesssim |\sin\theta| \lesssim 20 \lambda_\tau   \, .
\end{eqnarray}
The constraint of the thermal equilibrium can be approximately written as
\begin{eqnarray}
\sin^2\theta  k^2 m_V^2  ({GeV}) \gtrsim 5.8 \times 10^{2} \frac{\pi^3 \sqrt{g_\ast} ~ m_t}{\zeta (3) m_{\rm Pl}^{}} (\frac{m_V^{}}{20})^4 \approx 2.3 \times 10^{-12} (\frac{m_V^{}}{20})^4 \,\, .
\end{eqnarray}
For the vectorial WIMPs of concern, this thermal equilibrium constraint is below the neutrino background in DM direct detections, and the lower bound of the $\phi$ production cross section at $e^+ e^-$ collider is reduced by an order $10^{-5}$ factor compared with the case of scalar WIMPs.

\section{Conclusion and discussion}

The scalar and vectorial WIMPs have been studied in this article, with a new scalar $\phi$ as the mediator and the mass of $\phi$ being slightly below the WIMP mass. The dominant annihilation products of WIMPs are on-shell $\phi \phi$ pairs with $\phi$ mainly decaying into $\tau \bar \tau$, and the WIMP annihilations are phase space suppressed today. For masses of WIMPs in a range about 14 - 22 GeV, the annihilation cross section $\langle\sigma_{ann} v_r \rangle_0 \sim$  $1 \times 10^{-26}$ cm$^3/$s today can be obtained to meet the Galactic center GeV gamma-ray excess. Due to the nearly arbitrary small couplings between $\phi$ and SM particles, the WIMP-target nucleus SI scattering can be tolerant by the present stringent constraints of DM direct detections.

The upper limit of the $\phi$'s coupling to $\tau$ lepton is discussed with the constraints of WIMP annihilations, and the limit is tolerant by the muon $g - 2$ result. The scalar mediator-Higgs mixing angle $\theta$ should be small enough to keep $\tau \bar{\tau}$ dominant in the scalar $\phi$'s decay, and the upper limit of $\sin\theta$ from collider experiment is mild. The thermal equilibrium in the early universe sets an lower bound on the reaction rates of SM particles. Considering the $t \bar t$ contribution is dominant in SM particle reaction rates, we have derived an lower limit about the mixing angle $\theta$ and the coupling of the WIMP-$\phi$ trilinear term. For vectorial WIMPs, the reaction rate of $t \bar t \rightarrow$ WIMP pair is dramatically enhanced by the longitudinal polarization of vectorial WIMPs.

For the scalar WIMP-nucleon SI elastic scattering of concern, we obtain that the bound from the thermal equilibrium sets a minimum scattering cross section above the neutrino background. The present parameter spaces are set by the XENON1T result and the bound from thermal equilibrium, and the allowed region of the elastic scattering cross section is derived, with $\sigma_{\rm {el}} \sim$ $10^{-48} - 10^{-46}$ $\rm{cm}^2$. The range of the scattering cross section can be examined at the future DM ultimate direct detection experiments, such as LUX-ZEPLIN (LZ) \cite{Akerib:2015cja}, XENONnT \cite{Aprile:2015uzo} and DARWIN \cite{Aalbers:2016jon}. Thus, for WIMPs of concern, the future DM direct detections can give an answer about whether the scalar WIMP candidates exist or not. For vectorial WIMPs, the bound from the thermal equilibrium is below the neutrino background in direct detections, and this type WIMPs cannot be ruled out by the future DM ultimate direct detections.

The tests of the new sector at collider are as follows: i). The Higgs boson can decay into a WIMP pair, while this invisible decay is tiny and difficult to be explored at collider. ii). The decay channel $h \rightarrow \phi  \phi$ may leave traces at the future $e^+ e^-$ collider with the precise Higgs decay measurement, e.g. via the Higgs-strahlung process at $\sqrt{s} = 250 $ GeV, and a high luminosity of 500 fb$^{-1} $ $-$ 5 ab$^{-1}$ is needed. iii). For $\sqrt{s} < $ 400 GeV and above the threshold, the production of $\phi$ is mainly via the $\phi$ strahlung mechanism at $e^+ e^-$ collider. The signature of $\phi$ with the $\theta$ value near the upper limit $\sin^2\theta = 10^{-3}$ may appear at the future $e^+ e^-$ collider, and it is better to test the non-standard model $\phi$-like particle at a low center of mass energy collider with a high luminosity, e.g. $\sqrt{s} \sim$ 120 - 150 GeV with the energy above the $Z \phi$ production threshold and the luminosity up to about 200 fb$^{-1}$.

The future $e^+ e^-$ collider, such as the Circular Electron Positron Collider (CEPC) \cite{CEPC-Report}, the International Linear Collider (ILC) \cite{Baer:2013cma}, and FCC-ee(TLEP) \cite{Gomez-Ceballos:2013zzn}, may do the job to investigate the $h \rightarrow \phi  \phi$ decay and the production of $\phi$. We look forward to the future tests of the WIMPs of concern via DM indirect detections, DM direct detections and the hunt at collider.

\acknowledgments \vspace*{-3ex} Thank Xuewen Liu for useful discussions. This work was supported by the National Natural Science Foundation of China under Contract No. 11505144, and the Research Fund for the Doctoral Program of the Southwest University of Science and Technology under Contract No. 15zx7102.

\end{document}